\documentclass[draft,published,notoc]{JHEP} % 10pt is ignored!

\usepackage{epsfig}

\newcommand\fverb{\setbox\pippobox=\hbox\bgroup\verb}
\newcommand\fverbdo{\egroup\medskip\noindent%
			\fbox{\unhbox\pippobox}\ }
\newcommand\fverbit{\egroup\item[\fbox{\unhbox\pippobox}]}
\newbox\pippobox
\title{THE WEAK ISOSPIN AND THE GRAVITY}

\author{by Gunn A.Quznetsov\thanks{Special thanks to Prof. V.Dvoeglasov}\\ 
	E-mail: \email {quznets@geocities.com}}
\received{\today} 		%%
\accepted{\today}		%% These are for published papers.
\abstract{
The Clifford pentad of 4X4 complex matrices defines the currents of the
particles. The weak isospin transformation scatters the particle on two
components into the 2-dimensional space of the antidiagonal Clifford
matrices. The physics particles move in the 3-dimensional space of the
diagonal Clifford matrices. Such sectioning of the 5-dimensional space on
two subspace (2-dimensional and 3-dimensional) defines the Newtonian gravity
principle.}

\keywords{Classical Theory of Gravity, Standard Model}
\begin{document} 

\maketitle %%%%%%%%%% THIS IS IGNORED %%%%%%%%%%%

\section{Introduction}

In the Quantum Theory the fermion behavior is depicted by the spinor $\Psi .$
The probability current vector $\overrightarrow{j}$ components of this
fermion are the following:

\begin{equation}
j_x=\Psi ^{\dagger }\cdot \beta ^1\cdot \Psi ,j_y=\Psi ^{\dagger }\cdot
\beta ^2\cdot \Psi ,j_z=\Psi ^{\dagger }\cdot \beta ^3\cdot \Psi .
\label{w0}
\end{equation}

Here

\[
\beta ^1=\left[ 
\begin{array}{cccc}
0 & 1 & 0 & 0 \\ 
1 & 0 & 0 & 0 \\ 
0 & 0 & 0 & -1 \\ 
0 & 0 & -1 & 0
\end{array}
\right] ,\beta ^2=\left[ 
\begin{array}{cccc}
0 & -i & 0 & 0 \\ 
i & 0 & 0 & 0 \\ 
0 & 0 & 0 & i \\ 
0 & 0 & -i & 0
\end{array}
\right] ,\beta ^3=\left[ 
\begin{array}{cccc}
1 & 0 & 0 & 0 \\ 
0 & -1 & 0 & 0 \\ 
0 & 0 & -1 & 0 \\ 
0 & 0 & 0 & 1
\end{array}
\right]  
\]

are the members of the Clifford pentad, for which other members are the
following:

\[
\gamma ^0=\left[ 
\begin{array}{cccc}
0 & 0 & 1 & 0 \\ 
0 & 0 & 0 & 1 \\ 
1 & 0 & 0 & 0 \\ 
0 & 1 & 0 & 0
\end{array}\right], \beta ^4=\left[ 
\begin{array}{cccc}
0 & 0 & i & 0 \\ 
0 & 0 & 0 & i \\ 
-i & 0 & 0 & 0 \\ 
0 & -i & 0 & 0
\end{array}
\right] 
\]

Let this spinor be expressed in the following form:

\[
\Psi =\left| \Psi \right| \cdot \left[ 
\begin{array}{c}
\exp \left( i\cdot g\right) \cdot \cos \left( b\right) \cdot \cos \left(
a\right) \\ 
\exp \left( i\cdot d\right) \cdot \sin \left( b\right) \cdot \cos \left(
a\right) \\ 
\exp \left( i\cdot f\right) \cdot \cos \left( v\right) \cdot \sin \left(
a\right) \\ 
\exp \left( i\cdot q\right) \cdot \sin \left( v\right) \cdot \sin \left(
a\right)
\end{array}
\right] . 
\]

In this case the probability current vector $\overrightarrow{j}$ has got the
following components:

\begin{equation}
\begin{array}{c}
j_x=\left| \Psi \right| ^2\cdot \\ 
\cdot \left[ \cos ^2\left( a\right) \cdot \sin \left( 2\cdot b\right) \cdot
\cos \left( d-g\right) -\sin ^2\left( a\right) \cdot \sin \left( 2\cdot
v\right) \cdot \cos \left( q-f\right) \right] , \\ 
j_y=\left| \Psi \right| ^2\cdot \\ 
\cdot \left[ \cos ^2\left( a\right) \cdot \sin \left( 2\cdot b\right) \cdot
\sin \left( d-g\right) -\sin ^2\left( a\right) \cdot \sin \left( 2\cdot
v\right) \cdot \sin \left( q-f\right) \right] , \\ 
j_z=\left| \Psi \right| ^2\cdot \left[ \cos ^2\left( a\right) \cdot \cos
\left( 2\cdot b\right) -\sin ^2\left( a\right) \cdot \cos \left( 2\cdot
v\right) \right] .
\end{array}
\label{w1}
\end{equation}

.

If

\[
\rho =\Psi ^{\dagger }\cdot \Psi,  
\]

%%\begin{eqnarray}
%%\sum_{n=-\infty}^\infty x^n

then $\rho $ is the probability density, i.e. 
${\int \int \int_V }\rho \left( t\right) \cdot dV$ is the probability to find
the particle with the state function $\Psi $ in the domain $V$ of the
3-dimensional space at the time moment $t$. In this case, $\{\rho ,%
\overrightarrow{j}\}$ is the probability density $3+1$-vector.

If

\begin{equation}
\overrightarrow{j}=\rho \cdot \overrightarrow{u},  \label{w2}
\end{equation}

then $\overrightarrow{u}$ is the average velocity for this particle.

Let us denote:

\begin{equation}
J_0=\Psi ^{\dagger }\cdot \gamma ^0\cdot \Psi ,J_4=\Psi ^{\dagger }\cdot
\beta ^4\cdot \Psi ,J_0=\rho \cdot V_0,J_4=\rho \cdot V_4.  \label{w3}
\end{equation}

In this case:

\begin{equation}
\begin{array}{c}
V_0=\sin \left( 2\cdot a\right) \cdot \left[ \cos \left( b\right) \cdot \cos
\left( v\right) \cdot \cos \left( g-f\right) +\sin \left( b\right) \cdot
\sin \left( v\right) \cdot \cos \left( d-q\right) \right], \\ 
V_4=\sin \left( 2\cdot a\right) \cdot \left[ \cos \left( b\right) \cdot \cos
\left( v\right) \cdot \sin \left( g-f\right) +\sin \left( b\right) \cdot
\sin \left( v\right) \cdot \sin \left( d-q\right) \right] ;
\end{array}
\label{w4}
\end{equation}

and for every particle:

\begin{equation}
u_x+u_y+u_z+V_0^2+V_4^2=1.  \label{VV}
\end{equation}

For the left particle (for example, the left neutrino): $a=\frac \pi 2$,

\[
\Psi _L=\left| \Psi _L\right| \cdot \left[ 
\begin{array}{c}
0 \\ 
0 \\ 
\exp \left( i\cdot f\right) \cdot \cos \left( v\right) \\ 
\exp \left( i\cdot q\right) \cdot \sin \left( v\right)
\end{array}
\right] 
\]

and from ($\ref{w1}$), and ($\ref{w2}$): $u_x+u_y+u_z=1$.Hence, the left
particle velocity equals $1$; hence, the mass of the left particle equals to
zero.

Let $U$ be the weak global isospin (SU(2)) transformation with the
eigenvalues $\exp \left( i\cdot \lambda \right) $.

In this case for this transformation eigenvector $\psi $:

\[
U\psi =\left| \psi \right| \cdot \left[ 
\begin{array}{c}
\exp \left( i\cdot g\right) \cdot \cos \left( b\right) \cdot \cos \left(
a\right) \\ 
\exp \left( i\cdot d\right) \cdot \sin \left( b\right) \cdot \cos \left(
a\right) \\ 
\exp \left( i\cdot \lambda \right) \cdot \exp \left( i\cdot f\right) \cdot
\cos \left( v\right) \cdot \sin \left( a\right) \\ 
\exp \left( i\cdot \lambda \right) \cdot \exp \left( i\cdot q\right) \cdot
\sin \left( v\right) \cdot \sin \left( a\right)
\end{array}
\right] 
\]

and for $1\leq \mu \leq 3$ from (\ref{w1}):

\begin{equation}
\left( U\psi \right) ^{\dagger }\cdot \beta ^\mu \cdot \left( U\psi \right)
=\psi ^{\dagger }\cdot \beta ^\mu \cdot \psi ,  \label{w5}
\end{equation}

but for $\mu =0$ and $\mu =4$ from (\ref{w4}):

\begin{equation}
\begin{array}{c}
\begin{array}{c}
\psi ^{\dagger }\cdot \gamma ^0\cdot \psi =\left| \psi \right| ^2\cdot \sin
\left( 2\cdot a\right) \cdot \\ 
\left[ \cos \left( b\right) \cdot \cos \left( v\right) \cdot \cos \left(
g-f-\lambda \right) +\sin \left( b\right) \cdot \sin \left( v\right) \cdot
\cos \left( d-q-\lambda \right) \right],
\end{array}
\\ 
\begin{array}{c}
\psi ^{\dagger }\cdot \beta ^4\cdot \psi =\left| \psi \right| ^2\cdot \sin
\left( 2\cdot a\right) \cdot \\ 
\left[ \cos \left( b\right) \cdot \cos \left( v\right) \cdot \sin \left(
g-f-\lambda \right) +\sin \left( b\right) \cdot \sin \left( v\right) \cdot
\sin \left( d-q-\lambda \right) \right];
\end{array}
\end{array}
\label{w6}
\end{equation}

\section{ THE WEAK ISOSPIN SPACE}

In the weak isospin theory we have got the following entities ({Global
Symmetries, Standard Model}):

the right electron state vector $e_R$,

the left electron state vector $e_L$,

the electron state vector $e$ ($e=\left[ 
\begin{array}{c}
e_R \\ 
e_L
\end{array}
\right] $),

the left neutrino state vector $\nu _L$,

the zero vector right neutrino $\nu _R$.

the unitary $2\times 2$ matrix $U$ of the isospin transformation.($\det
\left( U\right) =1$) ({Gauge Symmetry}).

This matrix acts on the vectors of the kind:$\left[ 
\begin{array}{c}
\nu _L \\ 
e_L
\end{array}
\right] $.

Therefore, in this theory: if

\[
U=\left[ 
\begin{array}{cc}
u_{1,1} & u_{1,2} \\ 
u_{2,1} & u_{2,2}
\end{array}
\right] 
\]

then the matrix

\[
\left[ 
\begin{array}{cccc}
1 & 0 & 0 & 0 \\ 
0 & u_{1,1} & 0 & u_{1,2} \\ 
0 & 0 & 1 & 0 \\ 
0 & u_{2,1} & 0 & u_{2,2}
\end{array}
\right] 
\]

operates on the vector

\[
\left[ 
\begin{array}{c}
e_R \\ 
e_L \\ 
\nu _R \\ 
\nu _L
\end{array}
\right]. 
\]

Because $e_R$, $e_L$, $\nu _R$, $\nu _L$ are the two-component vectors then

\[
\left[ 
\begin{array}{c}
e_R \\ 
e_L \\ 
\nu _R \\ 
\nu _L
\end{array}
\right] is \left[ 
\begin{array}{c}
e_{R1} \\ 
e_{R2} \\ 
e_{L1} \\ 
e_{L2} \\ 
0 \\ 
0 \\ 
\nu _{L1} \\ 
\nu _{L2}
\end{array}
\right] 
\]

and

\[
\left[ 
\begin{array}{cccc}
1 & 0 & 0 & 0 \\ 
0 & u_{1,1} & 0 & u_{1,2} \\ 
0 & 0 & 1 & 0 \\ 
0 & u_{2,1} & 0 & u_{2,2}
\end{array}
\right] is  \underline{U}=\left[ 
\begin{array}{cccccccc}
1 & 0 & 0 & 0 & 0 & 0 & 0 & 0 \\ 
0 & 1 & 0 & 0 & 0 & 0 & 0 & 0 \\ 
0 & 0 & u_{1,1} & 0 & 0 & 0 & u_{1,2} & 0 \\ 
0 & 0 & 0 & u_{1,1} & 0 & 0 & 0 & u_{1,2} \\ 
0 & 0 & 0 & 0 & 1 & 0 & 0 & 0 \\ 
0 & 0 & 0 & 0 & 0 & 1 & 0 & 0 \\ 
0 & 0 & u_{2,1} & 0 & 0 & 0 & u_{2,2} & 0 \\ 
0 & 0 & 0 & u_{2,1} & 0 & 0 & 0 & u_{2,2}
\end{array}
\right]. 
\]

This matrix has got eight orthogonal normalized eigenvectors of kind:

\[
\underline{s_1}=\left[ 
\begin{array}{c}
1 \\ 
0 \\ 
0 \\ 
0 \\ 
0 \\ 
0 \\ 
0 \\ 
0
\end{array}
\right] ,\underline{s_2}=\left[ 
\begin{array}{c}
0 \\ 
1 \\ 
0 \\ 
0 \\ 
0 \\ 
0 \\ 
0 \\ 
0
\end{array}
\right] ,\underline{s_3}=\left[ 
\begin{array}{c}
0 \\ 
0 \\ 
\varpi \\ 
0 \\ 
0 \\ 
0 \\ 
\chi \\ 
0
\end{array}
\right] ,\underline{s_4}=\left[ 
\begin{array}{c}
0 \\ 
0 \\ 
0 \\ 
\chi ^{*} \\ 
0 \\ 
0 \\ 
0 \\ 
-\varpi ^{*}
\end{array}
\right] , 
\]

\[
\underline{s_5}=\left[ 
\begin{array}{c}
0 \\ 
0 \\ 
0 \\ 
0 \\ 
1 \\ 
0 \\ 
0 \\ 
0
\end{array}
\right] ,\underline{s_6}=\left[ 
\begin{array}{c}
0 \\ 
0 \\ 
0 \\ 
0 \\ 
0 \\ 
1 \\ 
0 \\ 
0
\end{array}
\right] ,\underline{s_7}=\left[ 
\begin{array}{c}
0 \\ 
0 \\ 
\chi ^{*} \\ 
0 \\ 
0 \\ 
0 \\ 
-\varpi ^{*} \\ 
0
\end{array}
\right] ,\underline{s_8}=\left[ 
\begin{array}{c}
0 \\ 
0 \\ 
0 \\ 
\varpi \\ 
0 \\ 
0 \\ 
0 \\ 
\chi
\end{array}
\right] . 
\]

The corresponding eigenvalues are: $1$, $1$, $\exp \left( i\cdot \lambda
\right) $, $\exp \left( i\cdot \lambda \right) $, $1$,$1$,

$\exp \left( -i\cdot \lambda \right) $, $\exp \left( -i\cdot \lambda \right) 
$.

These vectors constitute the orthogonal basis in this 8-dimensional space.

Let $\underline{\gamma ^0}=\left[ 
\begin{array}{cc}
\gamma ^0 & O \\ 
O & \gamma ^0
\end{array}
\right] $, if $O$ is zero $4\times 4$ matrix, and \underline{$\beta ^4$}$%
=\left[ 
\begin{array}{cc}
\beta ^4 & O \\ 
O & \beta ^4
\end{array}
\right] $.

The vectors $\left[ 
\begin{array}{c}
e_{R1} \\ 
e_{R2} \\ 
e_{L1} \\ 
e_{L2} \\ 
0 \\ 
0 \\ 
0 \\ 
0
\end{array}
\right] $, $\left[ 
\begin{array}{c}
e_{R1} \\ 
e_{R2} \\ 
0 \\ 
0 \\ 
0 \\ 
0 \\ 
0 \\ 
0
\end{array}
\right] $, $\left[ 
\begin{array}{c}
0 \\ 
0 \\ 
e_{L1} \\ 
e_{L2} \\ 
0 \\ 
0 \\ 
0 \\ 
0
\end{array}
\right] $ correspond to the state vectors $e$, $e_R$ and $e_L$ resp.

In this case (\ref{w3}) \underline{$e$}$^{\dagger }\cdot \underline{\gamma ^0%
}\cdot \underline{e}=J_{0e}$, \underline{$e$}$^{\dagger }\cdot \underline{%
\beta ^4}\cdot \underline{e}=J_{4e}$, $J_{0e}=\underline{e}^{\dagger }\cdot 
\underline{e}\cdot V_{0e}$, $J_{4e}=\underline{e}^{\dagger }\cdot \underline{%
e}\cdot V_{4e}$.

For the vector \underline{$e$} the numbers $k_3$, $k_4$, $k_7$, $k_8$ exist,
for which: \underline{$e$}$=(e_{R1}\cdot \underline{s_1}+e_{R2}\cdot 
\underline{s_2})+(k_3\cdot \underline{s_3}+k_4\cdot \underline{s_4}%
)+(k_7\cdot \underline{s_7}+k_8\cdot \underline{s_8})$.

Here \underline{$e_R$}$=(e_{R1}\cdot \underline{s_1}+e_{R2}\cdot \underline{%
s_2})$. If \underline{$e_{La}$}$=(k_3\cdot \underline{s_3}+k_4\cdot 
\underline{s_4})$ and \underline{$e_{Lb}$}$=(k_7\cdot \underline{s_7}%
+k_8\cdot \underline{s_8})$ then \underline{$U$}$\cdot \underline{e_{La}}%
=\exp \left( i\cdot \lambda \right) \cdot $\underline{$e_{La}$} and 
\underline{$U$}$\cdot \underline{e_{Lb}}=\exp \left( -i\cdot \lambda \right)
\cdot $\underline{$e_{Lb}$}.

Let for all $k$ ($1\leq k\leq 8$): \underline{$h_k$}$=\underline{\gamma ^0}%
\cdot $\underline{$s_k$}. The vectors \underline{$h_k$} constitute the
orthogonal basis, too. And the numbers $q_3$, $q_4$, $q_7$, $q_8$ exist, for
which: \underline{$e_R$}$=(q_3\cdot \underline{h_3}+q_4\cdot \underline{h_4}%
)+(q_7\cdot \underline{h_7}+q_8\cdot \underline{h_8})$.

Let \underline{$e_{Ra}$}$=(q_3\cdot \underline{h_3}+q_4\cdot \underline{h_4}%
) $, \underline{$e_{Rb}$}$=(q_7\cdot \underline{h_7}+q_8\cdot \underline{h_8}%
)$, \underline{$e_a$}$=\underline{e_{Ra}}+$\underline{$e_{La}$} and 
\underline{$e_b$}$=\underline{e_{Rb}}+$\underline{$e_{Lb}$}.

Let \underline{$e_a$}$^{\dagger }\cdot \underline{\gamma ^0}\cdot \underline{%
e_a}=J_{0a}$, \underline{$e_a$}$^{\dagger }\cdot \underline{\beta ^4}\cdot 
\underline{e_a}=J_{4a}$, $J_{0a}=\underline{e_a}^{\dagger }\cdot \underline{%
e_a}\cdot V_{0a}$, $J_{4a}=\underline{e_a}^{\dagger }\cdot \underline{e_a}%
\cdot V_{4a}$,

\underline{$e_b$}$^{\dagger }\cdot \underline{\gamma ^0}\cdot \underline{e_b}%
=J_{0b}$, \underline{$e_b$}$^{\dagger }\cdot \underline{\beta ^4}\cdot 
\underline{e_b}=J_{4b}$, $J_{0b}=\underline{e_b}^{\dagger }\cdot \underline{%
e_b}\cdot V_{0b}$, $J_{4b}=\underline{e_b}^{\dagger }\cdot \underline{e_b}%
\cdot V_{4b}$.

In this case: $J_0=J_{0a}+J_{0b}$, $J_4=J_{4a}+J_{4b}$.

Let $\left( \underline{U}\cdot \underline{e_a}\right) ^{\dagger }\cdot 
\underline{\gamma ^0}\cdot \left( \underline{U}\cdot \underline{e_a}\right)
=J_{0a}^{\prime }$, $\left( \underline{U}\cdot \underline{e_a}\right)
^{\dagger }\cdot \underline{\beta ^4}\cdot \left( \underline{U}%
\cdot \underline{e_a}\right) =J_{4a}^{\prime }$, $J_{0a}^{\prime }=\left( 
\underline{U}\cdot \underline{e_a}\right) ^{\dagger }\cdot \left( \underline{%
U}\cdot \underline{e_a}\right) \cdot V_{0a}^{\prime }$, $J_{4a}^{\prime
}=\left( \underline{U}\cdot \underline{e_a}\right) ^{\dagger }\cdot \left( 
\underline{U}\cdot \underline{e_a}\right) \cdot V_{4a}^{\prime }$,

$\left( \underline{U}\cdot \underline{e_b}\right) ^{\dagger }\cdot 
\underline{\gamma ^0}\cdot \left( \underline{U}\cdot \underline{e_b}\right)
=J_{0b}^{\prime }$, $\left( \underline{U}\cdot \underline{e_b}\right)
^{\dagger }\cdot \underline{\beta ^4}\cdot \left( \underline{U}\cdot 
\underline{e_b}\right) =J_{4b}^{\prime }$, $J_{0b}^{\prime }=\left( 
\underline{U}\cdot \underline{e_b}\right) ^{\dagger }\cdot \left( \underline{%
U}\cdot \underline{e_b}\right) \cdot V_{0b}^{\prime }$, $J_{4b}^{\prime
}=\left( \underline{U}\cdot \underline{e_b}\right) ^{\dagger }\cdot \left( 
\underline{U}\cdot \underline{e_b}\right) \cdot V_{4b}^{\prime }$.

In this case from (\ref{w6}):

\[
\begin{array}{c}
V_{0a}^{\prime }=V_{0a}\cdot \cos \left( \lambda \right) +V_{4a}\cdot \sin
\left( \lambda \right) , \\ 
V_{4a}^{\prime }=V_{4a}\cdot \cos \left( \lambda \right) -V_{0a}\cdot \sin
\left( \lambda \right) ; \\ 
V_{0b}^{\prime }=V_{0b}\cdot \cos \left( \lambda \right) -V_{4b}\cdot \sin
\left( \lambda \right) , \\ 
V_{4b}^{\prime }=V_{4b}\cdot \cos \left( \lambda \right) +V_{0b}\cdot \sin
\left( \lambda \right).
\end{array}
\]

Hence, every isospin transformation divides a electron on two components,
which scatter on the angle $2\cdot \lambda $ in the space of ($J_0$, $J_4$).

These components are indiscernible in the space of ($j_x$, $j_y$, $j_z$) (%
\ref{w5}).

Let $o$ be the $2\times 2$ zeros matrix. Let the $4\times 4$ matrices of
kind:

\[
\left[ 
\begin{array}{cc}
P & o \\ 
o & S
\end{array}
\right] 
\]

be denoted as the diagonal matrices, and

\[
\left[ 
\begin{array}{cc}
o & P \\ 
S & o
\end{array}
\right] 
\]

be denoted as the antidiogonal matrices.

Three diagonal members ($\beta ^1$, $\beta ^2$, $\beta ^3$) of the Clifford
pentad define the 3-dimensional space $\Re $, in which $u_x$, $u_y$ , $u_z$
are located. The physics objects move in this space. Two antidiogonal
members ($\gamma ^0$, $\beta ^4$) of this pentad define the 2-dimensional
space $\grave A$, in which $V_0$ and $V_4$ are located. The weak isospin
transformation acts in this space.

\section{GRAVITY}

Let $x$ be the particle average coordinate in $\Re $, and ${\bf m}$ be the
average coordinate of this particle in $\grave A$. Let $x+i\cdot {\bf m}$ be
denoted as the complex coordinate of this particle.

From (\ref{VV}) this particle average velocity, which proportional to $%
x+i\cdot {\bf m}$, is:

\[
v=\frac{x+i\cdot {\bf m}}{\sqrt{\left( x^2+{\bf m}^2\right) }}. 
\]

$\left| v\right| =1$, but for the acceleration:

\[
a=\frac{dv}{dt}=\frac{dv}{dx}\cdot v=-i\cdot {\bf m\cdot }\left( \frac{%
x+i\cdot {\bf m}}{x^2+{\bf m}^2}\right) ^2. 
\]

And if ${\bf m}\ll x$, then

\[
\left| a\right| \simeq \frac{{\bf m}}{x^2}. 
\]

This is very much reminds the Newtonian gravity principle (Classical
Theories of Gravity).

\end{document}